\documentclass{elsart5p}

\usepackage{graphicx}

\usepackage{amssymb}

\begin{document}

\begin{frontmatter}



\title{
Photo-production of neutral kaons on $^{12}$C~ in the threshold region
}


\author[tohoku]{T.~Watanabe\corauthref{tt}\thanksref{trgifu}},
\ead{ takaomi@phys.ed.gifu-u.ac.jp }
\corauth[tt]{Corresponding author, Tel: +81-58-293-2245; 
Fax: +81-58-293-2245}
\thanks[trgifu]{Present address: Gifu University, Gifu, 501-1193, Japan }
\author[npi]{P.~Byd\v{z}ovsk\'y},
\author[tohoku]{K.~Dobashi},
\author[akita]{S.~Endo},
\author[tohoku]{Y.~Fujii},
\author[tohoku]{O.~Hashimoto},
\author[lns]{T.~Ishikawa},
\author[tohoku]{K.~Itoh},
\author[tohoku]{H.~Kanda},
\author[tohoku]{M.~Katoh},
\author[lns]{T.~Kinoshita},
\author[ichinoseki]{O.~Konno},
\author[tohoku]{K.~Maeda},
\author[tohoku]{A.~Matsumura},
\author[lns]{F.~Miyahara},
\author[tohoku]{H.~Miyase},
\author[tohoku]{T.~Miyoshi},
\author[tohoku]{K.~Mizunuma},
\author[tohoku]{Y.~Miura},
\author[tohoku]{S.~N.~Nakamura},
\author[tohoku]{H.~Nomura},
\author[tohoku]{Y.~Okayasu}, 
\author[tohoku]{T.~Osaka},
\author[tohoku]{M.~Oyamada},
\author[akita]{A.~Sasaki},
\author[akita]{T.~Satoh},
\author[lns]{H.~Shimizu},
\author[npi]{M.~Sotona},
\author[tohoku]{T.~Takahashi\thanksref{trkek}},
\thanks[trkek]{Present address: Institute of Particle and Nuclear Studies,
High Energy Accelerator Organization (KEK), Tukuba, 305-0801, Japan}
\author[lns]{T.~Tamae},
\author[tohoku]{H.~Tamura},
\author[lns]{T.~Terasawa},
\author[tohoku]{H.~Tsubota},
\author[tohoku]{K.~Tsukada},
\author[tohoku]{M.~Ukai},
\author[tohoku]{M.~Wakamatsu}
\author[tohoku]{H.~Yamauchi},
\author[lns]{H.~Yamazaki},

\address[tohoku]{
Department of Physics, Tohoku University, Sendai, 980-8578, Japan
}
\address[npi]{
Nuclear Physics Institute, 25068, $\check{R}$e$\check{z}$, Czech Republic 
}
\address[akita]{
Department of Electrical and Electronic Engineering, 
Akita University, Akita, 010-8502, Japan 
}
\address[lns]{
Laboratory of Nuclear Science, Tohoku University, Sendai, 982-0826, Japan
}
\address[ichinoseki]{
Department of Electrical Engineering, Ichinoseki National College 
of Technology, Ichinoseki, 021-8511, Japan
}

\begin{abstract}

Kaon photo-production process on $^{12}$C has been
studied by measuring neutral kaons
in a photon energy range of 0.8$-$1.1 GeV.
Neutral kaons were identified by the invariant mass constructed from two charged pions
emitted in the $K^{0}_{S}\rightarrow\pi^{+}\pi^{-}$ decay channel.
The differential cross sections as well as the integrated ones
in the threshold photon energy region were obtained.
The obtained momentum spectra
were compared with a Spectator model calculation using
elementary amplitudes of kaon photo-production given by recent 
isobar models.
Present result provides, for the first time, 
the information on $n(\gamma,K^{0})\Lambda$
reaction which is expected to play an important role to construct models 
for strangeness production by the electromagnetic interaction.
Experimental results show
that cross section of $^{12}{\rm C}(\gamma,K^0)$
is of the same order to that of $^{12}{\rm C}(\gamma,K^+)$
and
suggest that slightly backward $K^0$ angular distribution is favored
in the $\gamma n\rightarrow K^0\Lambda$ process.
\end{abstract}

\end{frontmatter}

\section{Introduction}
\label{intro}

Strangeness photo-production process near the threshold
attracts strong interests in the context of the baryon resonances 
which couple with kaon/hyperon channel.
Integrated and differential cross sections as well as hyperon polarizations
in the $\gamma + p \rightarrow K^{+} + \Lambda$ and
$\gamma + p \rightarrow K^{+} + \Sigma^{0}$ reactions were
measured with high statistics by the SAPHIR collaboration~\cite{saphir98}.
Theoretical analysis in the framework of the isobar model showed
that the observed structure around 1900 MeV can be explained well
by including a new $D_{13}$ resonance of 1895 MeV~\cite{kmaid}.
It was also pointed out
that the values of the extracted resonance parameters are strongly influenced
by the treatment of background processes, namely, the choice of
the meson and hyperon resonances in t- and u-channels,
and the adopted recipes for the phenomenological 
hadronic form factors~\cite{janssen}.
The recent CLAS~\cite{McNabb}\cite{Bradford} and SAPHIR~\cite{saphir04} data
show a more pronounced structure around 1900 MeV than the old SAPHIR data.
Attempts to fit the models to these data have been made~\cite{bydzovsky},
however, new experimental data on such as spin observables~
\cite{McNabb}\cite{LEPS} 
and other isospin channels are necessary in order to properly construct the
theoretical models and to obtain clear conclusions on the resonances.
Strangeness photoproduction processes are expected to shine rich characteristics
of the hadron structures and the strangeness-involved reactions.

From the viewpoint of the isobar model, 
the reaction channel, $\gamma + n \rightarrow K^{0} + \Lambda$, 
in the threshold region has the following unique features;
(1) The t-channel Born term does not contribute.
(2) It is a mirror reaction to $\gamma + p \rightarrow K^{+} + \Lambda$.
A coupling constant, $g_{K\Sigma N}$, changes its sign because of the isospin
symmetry, $g_{K^{0}\Sigma^{0}n} = -g_{K^{+}\Sigma^{0}p}$,
resulting in the different interference effect.
This also holds for the exchange terms of isovector hyperon resonance 
in the u-channel.
(3) Contribution of higher mass resonances are suppressed 
in the threshold region.

It was also suggested that the angular distribution of neutral kaons 
could be backward peaked~\cite{Bennhold92}.  
Moreover, the amplitude of the $K^{0}\Lambda$ channel,
being sensitive only to couplings of the neutral particles,
is supposed to better differentiate diagrams in the isobar models. 
Thus, the neutral reaction channel is expected to play
a unique role in the investigation of the photo-production of strangeness,
which cannot be revealed only by measuring charged kaon channels.
 
In this paper, we report the first measurement of the $(\gamma, K^0)$ reaction
on a nucleus ($^{12}{\rm C}$) using a tagged photon beam. 
The integrated and differential cross sections are presented 
as a function of the incident photon energy, $ E_{\gamma}$.
The experimental results are compared with predictions of 
a Spectator model in which the elementary amplitudes of
the $\gamma + n \rightarrow K^{0} + \Lambda$ given by isobar models
were assumed.

\section{Experiment}

The experiment was performed using the internal tagged photon beam 
facility~\cite{stbtagger} at Laboratory of Nuclear Science, 
Tohoku University (~LNS~).

The 1.2 GeV electron beam
in the STretcher Booster ring (~STB ring~) produces
bremsstrahlung photons at the internal target 
of a thin carbon fiber of 15 $\mu$m $\phi$.
The photons were tagged over the energy range 
of 0.8 $\leq E_{\gamma} \leq$ 1.1~GeV with
$\Delta E_{\gamma} = \pm 10~{\rm MeV} $ 
by bremsstrahlung-recoil electrons, momenta of which
were analyzed by a bending magnet of the STB ring.
The average tagged photon rate was $ \sim 2.5 \times 10^6~\gamma/{\rm s}$.
The photons irradiated a ${\rm 2.1~g/cm^2}$ thick graphite target
( natural )
which was placed at the center of the Neutral Kaon Spectrometer (~NKS~)
as illustrated in Fig. 1.

\begin{figure}
 \begin{center}
  \includegraphics[scale=0.35]{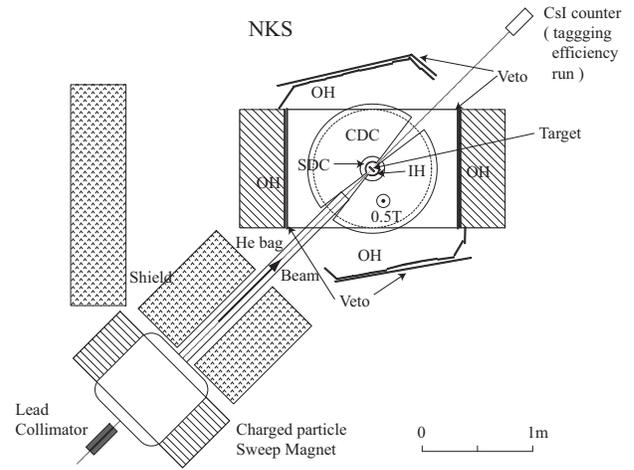}
 \end{center}
 \caption{
 Plan view of the Neutral Kaon Spectrometer (~NKS~)
 for the $^{12}$C($\gamma,K^0$) experiment.
 } 
 \label{setup}
\end{figure}

Charged particles produced
at the extraction window of the accelerator ring
and the collimator for eliminating the beam halo
were swept out by a sweep magnet followed 
by concrete shields.
The beam line from the sweep magnet to the target was filled with 
a helium bag to suppress the conversion positron-electron pairs which is
one of the major source of trigger backgrounds.
As shown in Fig.~\ref{setup}, a CsI(pure) counter
was placed downstream of the NKS
to measure the photon tagging efficiency
in the separate runs
with low intensity beams.
The averaged efficiency was measured to be 78$\pm$1 \%.

Neutral kaons were measured by detecting positive and negative 
pions in coincidence,
which were emitted in the $K^{0}_{S}\rightarrow\pi^{+}\pi^{-}$ decay channel.
The NKS was based on the TAGX spectrometer~\cite{tagx}, originally built
at Electron Synchrotron Laboratory of the Institute for Nuclear Study (INS-ES),
and was re-assembled at LNS.
It comprises a dipole magnet of 107 cm $\phi$ pole and 60 cm gap, 
straw (~SDC~) and cylindrical (~CDC~) drift chambers 
in the magnetic field of 0.5 Tesla,
inner (~IH~) and outer (~OH~) plastic scintillator hodoscopes, 
and veto plastic counters (~Veto~) in the beam plane,
covering 25\% of whole solid angle.

Since background triggers due to the conversion process
which occurred along the beam line were dominant,
Veto of 4 cm high
were installed in the mid-plane behind the OH counter arrays
to suppress them.
The trigger rates were typically 100 -- 200 Hz, but more than half was 
background triggers.
These background events were due to accidental coincidence between
the NKS trigger and the tagger trigger, and were easily rejected
by offline analysis.  

\section{Analysis} 

Horizontal momentum of a charged particle was reconstructed by 
the hit position information of SDC and CDC 
using the field map calculated by a 3D magnetic field program, TOSCA.
The vertical direction of the trajectory was calculated 
by approximate height at the target, which was assumed the same 
as that of the beams, and hit position at OH, which was obtained 
from the time difference between two PMT signals.
Time of flight was measured between the IH and the OH, 
where the distance is about 1 m. The time resolution was 0.6 ns (~rms~)
and was good enough to separate pions from protons below 0.7 GeV/$c$,
which was the maximum momentum of the decay pions in the present kinematics.
The $e^{+}e^{-}$ events were removed by rejecting the events of which
the vertex position was upstream the target.

\begin{figure}
\begin{center}
 \includegraphics[width=\linewidth]{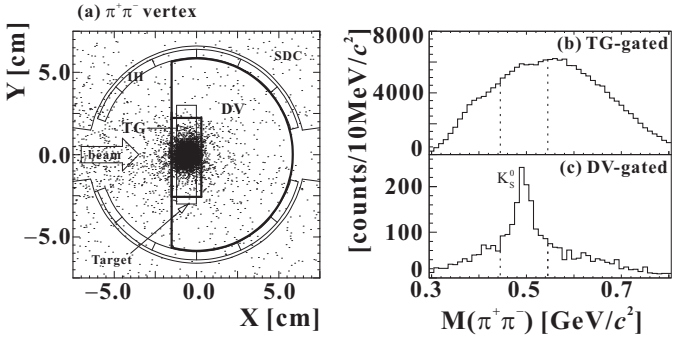}
\end{center}
\caption{
(a) A vertex distribution of $\pi^{+}\pi^{-}$ events.
The figure are a top view of the target area.
Beam comes from left to right (X-axis). 
The Y-axis means horizontal direction perpendicular to the beam.
Events come mainly from the target region denoted by TG.
(b) An invariant mass spectrum of $\pi^{+}\pi^{-}$ events
gated that the vertex is in the target region (~TG-gated~). 
(c) An invariant mass spectrum of $\pi^{+}\pi^{-}$ events
gated that the vertex is outside the target denoted by DV. 
The peak around $M=493~\rm{MeV}$ is identified as $K^{0}_{S}$.
}
\label{fig2}
\end{figure}

Fig.~\ref{fig2}(a) shows the vertex point distribution
of $\pi^{+}\pi^{-}$ events.
An opening angle (~$\eta$~) cut, $\cos\eta > -0.8$,  was applied in order to
keep a good vertex resolution, which was 1.7 mm.
Almost all of the events were produced in the target denoted 
as TG in Fig. 2(a).
They come from background processes such as a multi-pion production,
$N^{*}$, and $\rho^{0}$.
The invariant mass spectrum for these events
does not show any peak structure as shown in Fig.~\ref{fig2}(b).
On the other hand, the invariant mass spectrum for those events
whose vertex points are in the decay volume region (~DV~)
shows a peak structure at the $K^{0}$ mass as demonstrated 
in Fig.~\ref{fig2}(c).
Since $K^{0}_{S}$ has relatively long lifetime of $c\tau$=2.68 cm and
decays in flight, the $K^{0}$ events are enhanced by selecting
the vertex points outside the target.

\begin{figure}
 \begin{center}
  \includegraphics[width=0.8\linewidth]{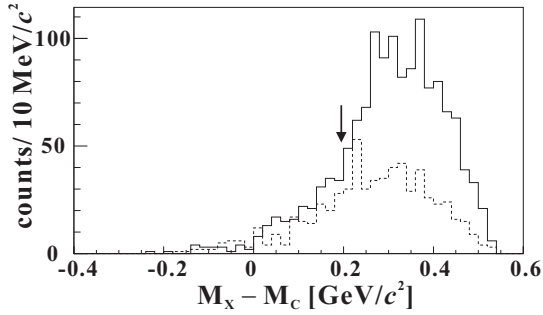}
 \end{center}
 \caption{
 Missing mass spectrum of $\gamma{\rm C}\rightarrow K^0 {\rm X}$ (solid).
 Background is estimated from invariant mass region of
 $0.395 < {\rm M}_{\pi^+\pi^-} < 0.445$ and
 $0.545 < {\rm M}_{\pi^+\pi^-} < 0.595$ (dashed).
 Arrow shows threshold of the quasi-free production.
}
\label{figMissMassC}
\end{figure}

In addition to the vertex position cut, the kinematical consistency
between the vertex position and the two-body momentum direction
was required
in order to further reject the mis-reconstructed events.
Missing mass spectrum after applying all selection
is shown in fig.~\ref{figMissMassC}.
It shows measured $K^0$s are produced
in the quasi-free kinematical region.

\begin{figure}
 \begin{center}
  \includegraphics[width=0.8\linewidth]{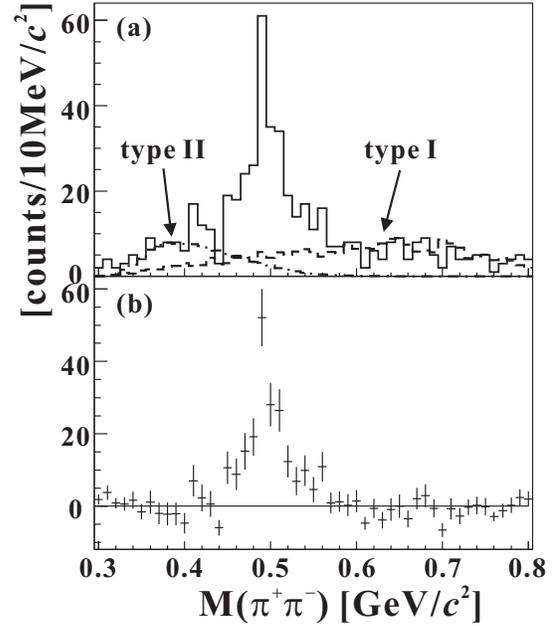}
 \caption{
  Invariant mass spectra without (a) and with (b)
  the estimated background subtraction.
  In (a), solid, dashed and dot-dashed lines are
  raw, type(I) background and type(II) background, respectively.
  The data is of $1.05<E_\gamma<1.10$ and $0.9<\cos\theta_{K0}<1.0$.
  }
 \label{FigBackgroundShapeIM}
 \end{center}
\end{figure}

\begin{figure}
 \begin{center}
  \includegraphics[width=0.8\linewidth]{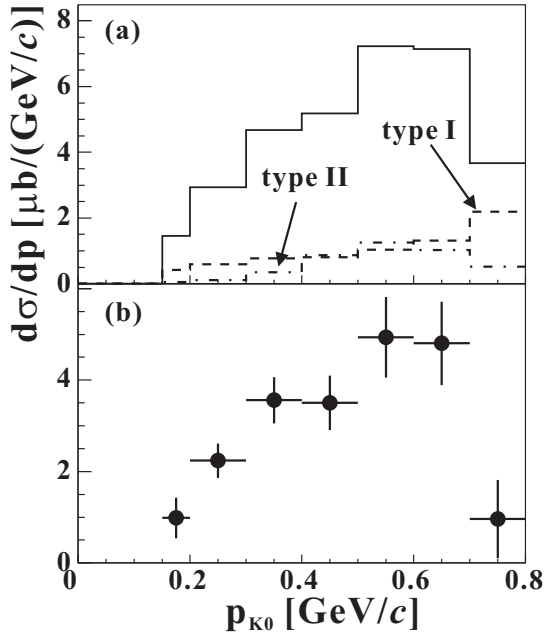}
  \caption{
  Momentum spectra without (a) and with (b)
  the estimated background subtraction.
  In (a), solid, dashed and dot-dashed lines are
  raw, type(I) background and type(II) background, respectively.
  The data is of $1.00<E_\gamma<1.10$ and $0.9<\cos\theta_{K0}<1.0$.
  }
 \label{FigBackgroundShapeP}
 \end{center}
\end{figure}

Momentum spectra of neutral kaons were obtained by subtracting
background events assuming they have the following two origins;
(I) reaction processes other than $K^{0}$ production
such as $\rho$ production,
of which vertex should be reconstructed in the TG region
but was reconstructed in the DV region
due to the limited vertex resolution.
(II) combinational background of $\pi^+\pi^-$,
such as $\pi^+$ from $K^0_S$ decay
and $\pi^-$ from $\Lambda$ decay.
The shapes of the background in the invariant mass and momentum distribution
were calculated from the experimental data for (I)
and from a Monte Calro simulation using Geant4~\cite{geant4} for (II),
respectively.
The ratio of the background to the $K^{0}_{S}$
was obtained by fitting the invariant mass spectra
at each photon-energy.
In fig.~\ref{FigBackgroundShapeIM}(a),
the background shapes in the invariant mass spectrum
are shown by dashed line for (I)
and dot-dashed line for (II).
Background subtracted invariant mass spectrum is shown
in fig.~\ref{FigBackgroundShapeIM}(b).
The $K^0$ momentum spectra before and after the background subtraction
are also shown in fig.~\ref{FigBackgroundShapeP}(a) and (b).

The spectrometer acceptance was estimated by a Geant4 simulation.
Analysis efficiencies were also estimated 
based on the simulation except for the tracking efficiency.
The tracking efficiency which depends on the intrinsic chamber efficiencies
and the track-finding algorithm was estimated to be 67\%
using pion events for left and right arms independently.
The numbers of tagged photons for each tagger segment
were counted with scalers.
The cluster hits and the analysis cuts
were corrected using the tagger trigger data taken in the same beam condition.
Total normalization error is estimated to be $^{+15}_{-7} $\%.
A large systematic error comes from strong position dependence of
the tracking efficiency due to high singles rate at forward angles.

\section{Results and discussion}

\begin{figure}
  \begin{center}
   \includegraphics[width=0.5\linewidth]{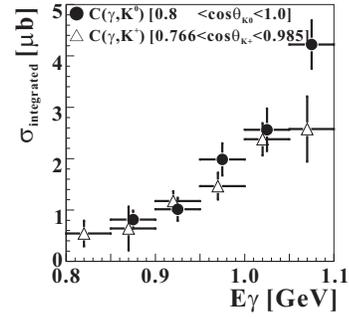}
  \end{center}
  \caption{
  The photon energy dependence of the integrated cross sections 
  for the kaon photo-production on $^{12}$C.
  Closed circles show the present data for $K^{0}$, integrated over the range,
  $0.8\:<\:\cos\theta_{K}^{\rm lab}\:<\:1.0$.
  Open triangles show $K^{+}$ data integrated
  from 10$^{\circ}$ to 40$^{\circ}$,
  which are taken from Ref.~\cite{yamazaki}.
  }
  \label{edepcs}
\end{figure}

The momentum- and angle-integrated cross sections
as a function of photon energy
for $K^{0}$ are plotted in Fig.~\ref{edepcs}, 
in which those for $K^{+}$~\cite{yamazaki} are overlaid.
The integrated momentum range was chosen 
so as to cover the kinematically allowed region
for the quasi-free process, namely,
from 0.15 to 0.70 GeV/$c$ for 0.9 $<$ $E_{\gamma}$ $<$ 1.0 GeV and
from 0.15 to 0.80 GeV/$c$ for 1.0 $<$ $E_{\gamma}$ $<$ 1.1 GeV,
where 0.15 GeV/$c$ is the detection threshold.
Angular ranges are practically the same to each other.
It can be concluded that the cross sections for $K^{0}$ and $K^{+}$ 
photo-production on $^{12}$C (T=0) are similar in magnitude,
suggesting that the elementary cross sections for 
$n(\gamma,K^{0})\Lambda$ and  $p(\gamma,K^{+})\Lambda$ are
of the same order.

\begin{figure*}
  \begin{center}
   \includegraphics[width=0.45\linewidth]{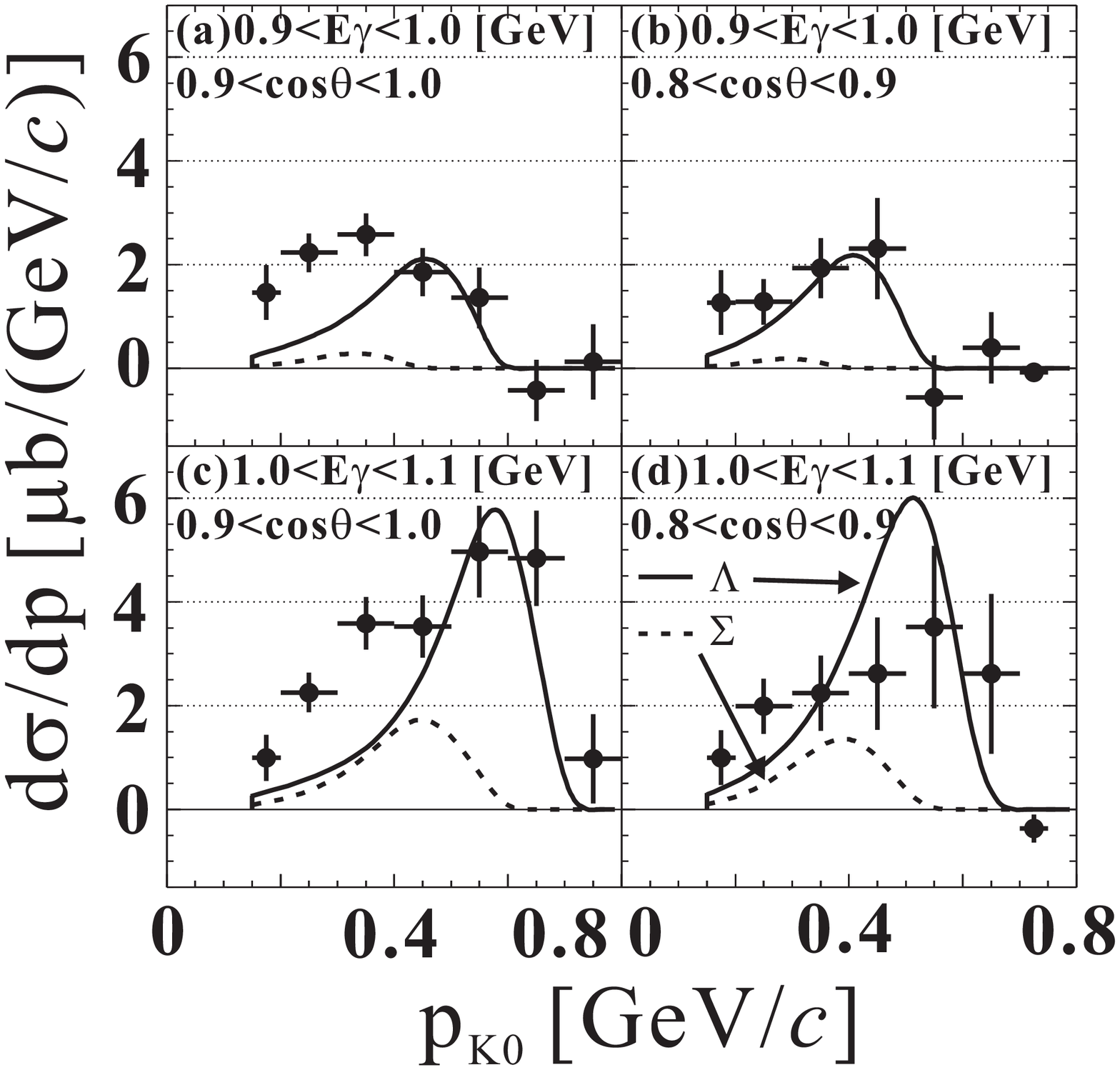}
   \includegraphics[width=0.45\linewidth]{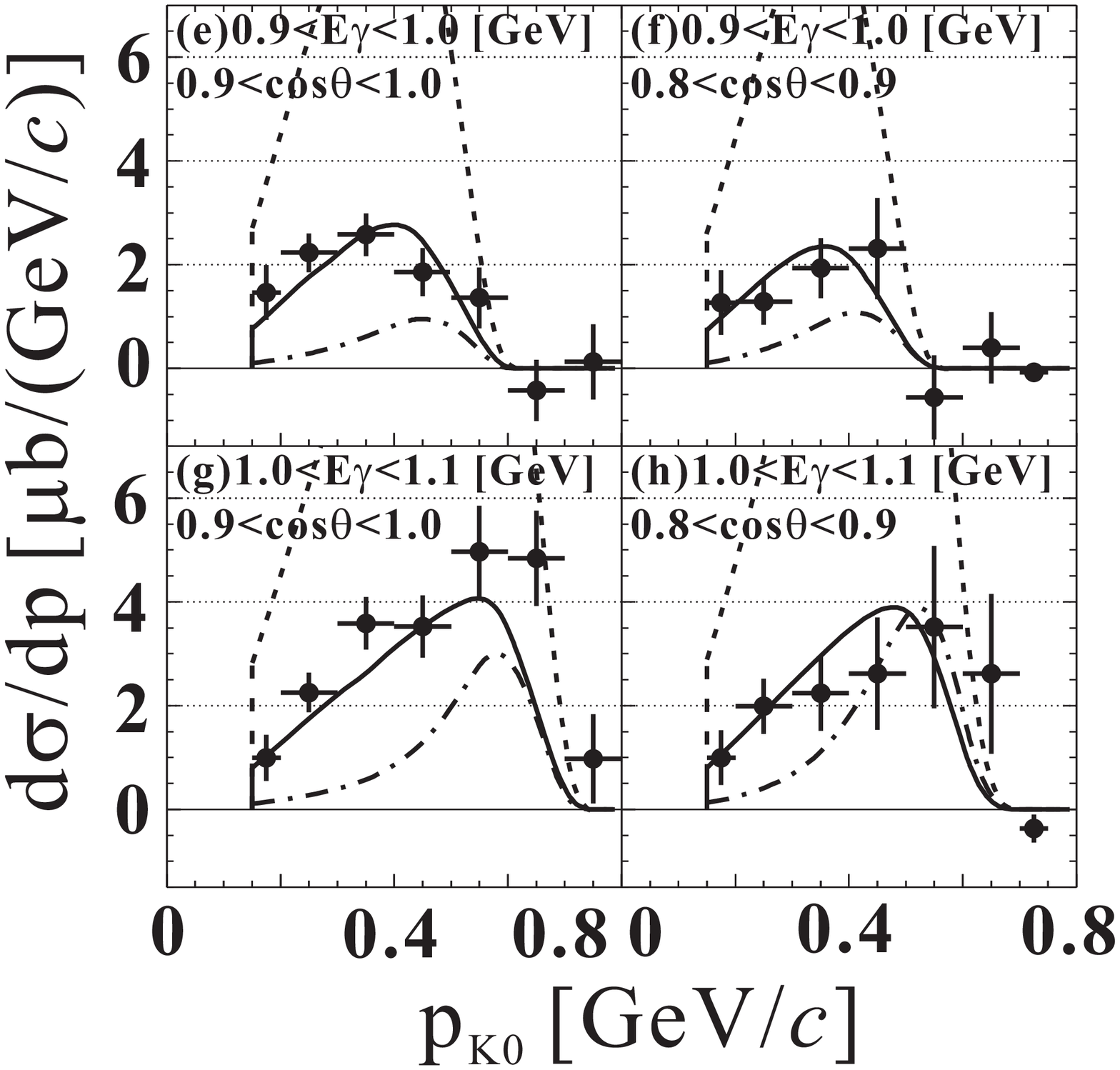}
  \end{center}
 \caption{
 Momentum spectra for $K^{0}$ with
 calculated cross sections using the elementary amplitudes of 
 kMAID\cite{kmaid} (a-d)
 and SLA\cite{SLA} (e-h) models.
 The photon energy and angular ranges are displayed in the figures.
 The error does not include the overall normalization error
 of $^{+15}_{-7}\%$.
 Solid and dashed curves in (a-d) show contributions from
 $n(\gamma,K^{0})\Lambda$ process and sum of $n(\gamma,K^{0})\Sigma^{0}$ and
 $p(\gamma,K^{0})\Sigma^{+}$ processes, respectively.
 Dashed, solid and dot-dashed lines in (e-h) show
 contribution of $n(\gamma,K^{0})\Lambda$ process
 with $r_{KK_1}=-0.447, -1.5$, and $-3.4$, respectively.
 }
\label{FigResult}
\end{figure*}

In Fig.~\ref{FigResult},
the obtained four momentum spectra of $K^{0}$ are shown 
for the two photon energy regions, 0.9 $<$ $E_{\gamma}$ $<$ 1.0 GeV and 
1.0 $<$ $E_{\gamma}$ $<$ 1.1 GeV,
and two $K^0$ angular ranges,
$0.8\:<\:\cos\theta_{K}^{\rm lab}\:<\:0.9$ and
$0.9\:<\:\cos\theta_{K}^{\rm lab}\:<\:1.0$.
The quoted errors include statistical ones
and the uncertainty of the acceptance.
In the high momentum region at the forward angle,
the errors becomes larger due to the smaller acceptance.

The obtained experimental results were compared with 
the theoretical calculations using a Spectator model. 
In this model the cross section of the quasi-free 
$\gamma + N \rightarrow K^{+,0} + Y$
was calculated as an incoherent sum of the elementary invariant amplitudes ($F$)
for the production on individual bound nucleons ($N$).
In the laboratory frame the cross section reads
\begin{eqnarray}
\frac{d^{2}\sigma}{d\Omega dp_{K}}  & = &
\frac{1}{(4\pi)^{2}}\int d\vec{p}_{N} \rho(\vec{p}_{N}) 
\frac{\left | F(s,t,\tilde{m}_{N}) \right |^{2}}
{4\:p_{\gamma}\cdot p_{N}} \nonumber \\ 
 & & \times \frac{{\rm p}_{K}^{2}}{E_{K} E_{Y}}
\delta(E_{\gamma}+\tilde{E}_{N}-E_{K}-E_{Y}) \nonumber \\ \nonumber
 & = &
\int d\vec{p}_{N}\rho(\vec{p}_{N})
\frac{d\sigma}{d\Omega}^{*}(W,\cos\theta^{*}_{\gamma K},\tilde{m}_{N})
\frac{{\rm p}^{*}_{\gamma}}{{\rm p}^{*}_{K}} 
\nonumber \\
 & & \times \frac{W^{2}}{p_{\gamma}\cdot p_{N}} 
 \frac{{\rm p}_{K}^{2}}{E_{K} E_{Y}}
\delta(E_{\gamma}+\tilde{E}_{N}-E_{K}-E_{Y}),
\end{eqnarray}
where $p$, $\vec{p}$, and p are the 4-momentum vector, 3-momentum
vector, and its magnitude, respectively. $\tilde{E}_N$ denotes the
energy of the bound nucleon, which is fixed by the
$\delta$-function, and $\tilde{m}_N$ the corresponding mass. 
W is the c.m. energy of the incoming photon and the bound nucleon. 
The variables denoted by * are those in the center-of-mass frame 
of the photon and nucleon. 
The momentum distribution of the bound nucleons,
$\rho(\vec{p}_N)$, is normalized to an effective proton
($Z_{eff}$) or neutron ($N_{eff}$) numbers of the target nucleus $A$. 
These phenomenological parameters mimic attenuation effects
of the initial photons, and final kaons and $\Lambda$ (the final
state interaction).
The value of 4.2 was assumed for both $Z_{eff}$ and $N_{eff}$,
which was deduced in the $^{12}$C($\gamma,K^+$) experiment in
the same energy region~\cite{yamazaki}. 
The momentum distribution of the nucleons was modeled by the Fermi 
gas model with $k_F=0.22$ GeV/$c$ for simplicity.

In Eq.(1) the conservation of energy and momentum was required
both in the complete (many-body) system and the elementary
(2-body) one
\begin{eqnarray}
p_{\gamma} + p_A & = & p_K + p_{\Lambda} + p_{A-1}, \\
p_{\gamma} + p_N & = & p_K + p_{\Lambda}.
\end{eqnarray}
Under these conditions the energy and momentum of the bound
nucleon are
\begin{eqnarray}
\tilde{E}_N & = & M_A - \sqrt{M_{A-1}^2 + {\rm p}_N^2}, \\
\vec{p}_N & = & -\vec{p}_{A-1},
\end{eqnarray}
where $M_A$ and $M_{A-1}$ are masses of the target and residual
nuclei, respectively.
Since the residual nucleus propagates on its mass-shell
the bound nucleon is off-shell and its energy therefore does not
correspond to the on-mass-shell value. 
For $K^0\Lambda$ production on $^{12}$C, 
the mass of the bound nucleon
decreases by 2\% at ${\rm p}_N$=0 GeV/$c$ and 5\% at
${\rm p}_N$=0.22 GeV/$c$.

This mass reduction makes the two-body c.m. energy lower. 
The kinematical area for integration is therefore smaller which
results in smaller cross sections than that from the on-mass-shell
approximation, $\tilde{E}_N = E_N = \sqrt{{\rm p}_N^2 +m_N^2}$.
Additional off-shell effects are given by the in-medium
modification of the elementary amplitude. 
In our approach the modification was done assuming the explicit 
dependence of the amplitude on the nucleon mass, 
$F(s,t,\tilde{m}_N)$, rather than adopting the completely 
relativistic description discussed in \cite{aw}.
This can be justified due to the small variations of $\tilde{m}_N$ 
in the kinematical region assumed here.

The elementary amplitude was evaluated using the Kaon-MAID (kMAID)
~\cite{kmaid} and Saclay-Lyon A (SLA)~\cite{SLA} models. 
Besides the Born terms these isobar models include  the 
$K^*(890)$ and $K_1(1270)$ diagrams which were shown to be important 
for a proper description of the data in the intermediate 
energy region~\cite{wjc}. 
In the SLA model four hyperon and one nucleon resonances are assumed 
in addition whereas only four nucleon resonances are included in kMAID. 
The structure in the hadronic vertices is
modeled by the hadronic form factors in kMAID but point like
hadrons are assumed in SLA.

In the isobar models, the ratios of the electromagnetic transition 
coupling constants between the charged and neutral particles have to 
be adjusted before the model is applied to the $K^0\Lambda$ channel. 
The ratios are known
for the processes involving a nucleon or $K^*$
but unknown for those with $K_1$. 
In kMAID the ratio, $r_{KK_1}=g(K_1^0K^0\gamma)/g(K_1^+K^+\gamma)$, 
was determined by simultaneously fitting the data of
$p(\gamma,K^+)\Lambda$, $p(\gamma,K^+)\Sigma^0$, and
$p(\gamma,K^0)\Sigma^+$ channels. The kMAID model provides,
therefore, predictions for both $K^0\Lambda$ and $K^0\Sigma^0$
channels. 
In SLA, however,
the ratio $r_{KK_1}$ is free
and has to be adjusted for $K^0$ production.

The results of calculations using the kMAID amplitudes are shown
in Fig.\ref{FigResult}(a-d), where the solid and dashed lines 
display contribution  from the $n(\gamma,K^0)\Lambda$ process and sum of
$n(\gamma,K^0)\Sigma^0$ and $p(\gamma,K^0)\Sigma^+$ processes, respectively. 
The calculations reproduce the present spectra reasonably well
both in the magnitude and shape. 
Although it is model dependent the results for the $K^0\Sigma^0$ and
$K^0\Sigma^+$ channels (dashed line) suggest that contribution of
these channels are relatively small at photon energy below 1 GeV.
At energies above 1 GeV, these processes become more important.

In Fig.~\ref{FigResult}(e-h),
the same data in Fig.~\ref{FigResult}(a-d)
are compared with the SLA amplitude.
Results for three values of the ratio $r_{KK_1}$, $-0.447$, $-1.5$,
and $-3.4$, are shown to demonstrate the sensitivity of the cross
sections to this parameter. 
The cross sections are greater with larger $r_{KK_1}$ values 
as shown in the figure. The best result
was achieved with the value $r_{KK_1}$= $-1.5$, whereas the value
$r_{KK_1}$= $-0.447$ used in the kMAID model overestimates the
present results. 
The value $r_{KK_1}$= $-3.4$, which gives very
similar predictions for the elementary cross sections as kMAID,
underestimates the presented experimental data. 
The SLA can also better describe the spectral shapes in the 
low $K^0$ momentum region with lower photon energy.

\begin{figure}[h]
 \begin{center}
  \includegraphics[width=0.8\linewidth]{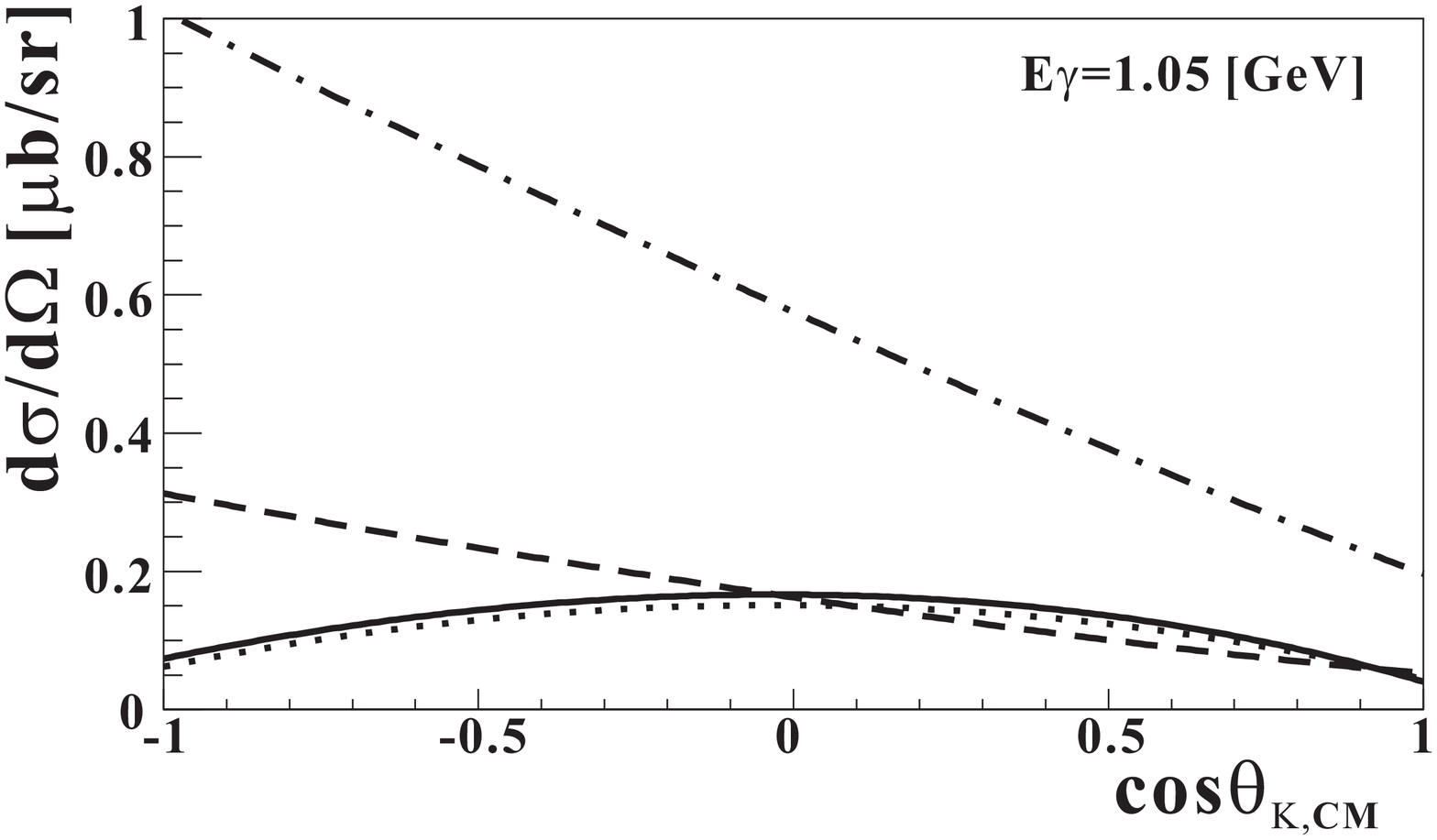}
 \end{center}
 \caption{Angular distribution
 for $n(\gamma,K^{0})\Lambda$ process
 at $E_{\gamma}=1.05$ GeV.
 Solid line shows kMaid model.
 Dot-dashed, dashed and dotted lines show
 SLA modesls of $r_{KK1} = -0.447$, $-1.5$ and $-3.4$, respectively.
 More details are in \cite{BydzovskySNP}.
 }
 \label{FigElemCS}
\end{figure}

Although $K^0$ angular distributions cannot be directly deduced
from the present data due to limited kinematical acceptance,
we evaluated them by comparing the measured momentum spectra
with those calculated by the models.
As shown in fig.\ref{FigElemCS},
kMAID and SLA with different $r_{KK_1}$ values
give different angular distribution
of the elementary $n(\gamma,K^0)\Lambda$ process
in the center-of-mass frame.
Converting the center-of-mass frame to those in the laboratory system,
the $K^0$ momentum spectra show quite different shapes,
backward $K^0$ in the center-of-mass frame
contributing more to those in the low momentum region
in the laboratory system
and forward $K^0$ to the higher momentum region
because of the Lorentz boost.
Since the SLA prediction with $r_{KK_1}=-1.5$,
which represents gentle backward angular distribution,
gives reasonable agreement with the present data,
it can be said without going into the detail of the models
that the present data possibly suggest
slightly backward angular distribution
for the elementary $\gamma n\rightarrow K^0\Lambda$ process,
though some part of the low momentum excess is
possibly due to the $\Sigma$ production.

\section{Summary}
We have, for the first time,
measured integrated and differential cross sections of
the $^{12}$C$(\gamma,K^{0})$ reaction at the photon energy below 1.1 GeV, 
identifying neutral kaons by reconstructing 
the $K^{0}_{S}\rightarrow\pi^{+}\pi^{-}$ decay.
It was found that the integrated cross section is almost the same 
in magnitude as that of $^{12}$C$(\gamma,K^{+})$.
Quasi-free spectra of the reaction were calculated
using the elementary amplitudes given
by the Kaon-MAID (kMAID) and Saclay-Lyon A (SLA) models
and were compared with the present experimental data. 
Both models explain the spectra in the threshold region reasonably well, 
though the SLA model can better account for the excess of the measured 
cross section in the $K^0$ low momentum region  
compared with the kMAID calculation.
It possibly suggests that $n(\gamma,K^{0})\Lambda$ reaction 
is more backward peaked in the center-of-mass frame.
The present data provide the first information on the unique 
strangeness photo-production in the neutral channel and 
demonstrate the importance of the $n(\gamma,K^{0})\Lambda$ reaction
for the investigation of the strangeness photo-production.
Further measurements with a deuterium target is underway.

\section*{Acknowledgment}
The authors thank the scientific and technical staffs of
LNS for the operation of the accelerator and 
experimental supports.
They are grateful to Prof. K.~Maruyama and Prof. H.~Okuno
for their help to move over the TAGX spectrometer 
from INS to Tohoku University.
They also thank Dr. T.~Mart for the useful discussion and offering
the program code.
This work is supported by Grant-In-Aid for Scientific
Research from The Ministry of Education of Japan,
Nos. 09304028, 12002001 and 16GS0201.
T.T. acknowledges the support from Grant-In-Aid for Scientific
Research from The Ministry of Education of Japan, No. 14740150.
P.B. and M.S. acknowledge support from Grant Agency of the Czech
Republic, Grant No. 202/05/2142.

\end{document}